\documentclass[journal=ancac3,manuscript=article]{achemso}

\usepackage{chemformula} 
\usepackage[T1]{fontenc} 

\usepackage{lineno}%
\usepackage{threeparttable} 
\usepackage{graphicx,amssymb,amsmath}
\usepackage{epstopdf}

\author{Xu Liu}
\altaffiliation{These authors contributed equally}
\affiliation[company]
{Two Pore Guys Inc., Santa Cruz, CA}
\author{Yuning Zhang}
\affiliation{Department of Physics, McGill University, Montreal, QC, Canada}
\altaffiliation{These authors contributed equally}
\author{Roland Nagel}
\affiliation[company]
{Two Pore Guys Inc., Santa Cruz, CA}
\author{Walter Reisner}
\affiliation{Department of Physics, McGill University, Montreal, QC, Canada}
\email{reisner@physics.mcgill.ca}
\author{\\ William B.~Dunbar}
\email{bill@twoporeguys.com}
\affiliation[company]
{Two Pore Guys Inc., Santa Cruz, CA}

\title[A dual nanopore device]
  {Controlling DNA Tug-of-War in a Dual Nanopore Device}


\begin{document}

%
%
%
%
%

\begin{abstract} 
Methods for reducing and directly controlling the speed of DNA through a nanopore are needed to enhance sensing performance for direct strand sequencing and detection/mapping of sequence-specific features.  We have created a method for reducing and controlling the speed of DNA that uses two independently controllable nanopores operated with an active control logic.   The pores are positioned sufficiently close to permit co-capture of a single DNA by both pores.  Once co-capture occurs, control logic turns on constant competing voltages at the pores leading to a ``tug-of-war'' whereby the molecule is pulled from both ends by opposing forces.  These forces exert both conformational and speed control over the co-captured molecule, removing folds and reducing the translocation rate.  When the voltages are tuned so that the electrophoretic force applied to both ends of the molecule comes into balance, the life-time of the tug-of-war state is limited purely by diffusive sliding of the DNA between the pores.  We are able to produce a tug-of-war state on 76.8\% of molecules that are captured with a maximum two-order of magnitude increase in average pore translocation time relative to the average time for single-pore translocation.  Moreover, we quantify the translocation slow-down as a function of voltage tuning and show that the slow-down is well described by a first passage analysis for a one-dimensional sub-diffusive process.  The ionic current of each nanopore provides an independent sensor that synchronously measures a different region of the same molecule, enabling sequential detection of physical labels, such as mono-streptavidin tags.  With advances in devices and control logic, future dual-pore applications include genome mapping and enzyme-free sequencing.
\end{abstract}


\newpage
     
     A nanopore is a nano-scale opening in a membrane that separates two fluidic chambers.  A voltage-clamp circuit supplies a voltage bias to capture and pass individual charged molecules, such as DNA, from one chamber to the other.  The same circuit measures the ionic current through the pore and detects each capture and passage ``event'' as a temporary attenuation in the current or ``blockade''.  This blockade signal provides an identifying signature for the molecules.  Nanopores have become a powerful tool for performing biological analysis, enabling label-free, low-cost sensing of single biomolecules using a purely electrical sensing approach.
     
       Controlling translocation is a key problem in the nanopore field:  voltages that promote capture and sufficient sensing signal also produce DNA passage rates that are 100 times too fast.\cite{Branton:2008fr}  For protein pores, the rate reduction problem has been addressed using enzymes to ratchet DNA through the pore in single-nucleotide steps,\cite{cherf2012automated, derrington2015subangstrom} a key ingredient in commercial nanopore sequencing technology.\cite{Jain:2015ik}  Solid-states devices, made out of conventional semiconductor materials such as silicon nitride, admit of more scalable fabrication\cite{miles2013} and potentially might have higher resolution.\cite{Lindsay2016}  Protein-pores are fundamentally limited by the number of bases that can fit in the pore at one time (e.g. groups of five bases, for the Oxford nanopore sequencer).\cite{Lindsay2016}  This resolution limit increases error-rate and necessitates complex algorithms to reconstitute the sequence.\cite{Jain2015}   A large number of solid-state sequencing schemes have been proposed to increase pore sensing resolution, for example implementing blockade based sensing with atomically thin films such as graphene or complementing blockade-based sensing with additional approaches such as tunneling electrodes and capacitance sensing.\cite{miles2013}  In addition, solid-state pores, due to their more flexible dimensions, are better suited for larger size analytes, such as dsDNA and proteins.  Applications include the detection of PNA (Peptide Nucleic Acid)-bound target sequences within dsDNA,\cite{Squires:2015ir} and the detection of bound proteins to signal the presence of modified bases.\cite{Shim:2015ha}
    
     Solid-state nanopores lack a translocation control mechanism that can slow-down the DNA while ensuring a high pore sensing voltage to generate ample SNR for robust feature detection.  In addition, for large diameter ($>5$\,nm) pores that enable efficient capture of long ($>10$\,kbp) DNA, the common occurrence of folds during translocation greatly reduces the number of usable events featuring a linear correspondence between translocation time and sequence position.\cite{keyserbarcode2} Inspired by the independence of sensors from actuators in the field of feedback control theory,\cite{Astrom:2010wa, Dunbar:2015vn} we have developed an active control approach using a dual nanopore device that exerts simultaneous conformational and speed control over the translocating single-molecules.  Our two-pore approach separates the dual roles of voltage: first, to capture and drive DNA through the pore, and second, to generate the sensing signal.   The two nanopores are positioned sufficiently close to permit co-capture of a single DNA and, using an-all insulator wafer-scale fabrication approach, fluidically configured to permit independent voltage control and current sensing.  Using control logic implemented by a Field-Programmable Gate Array (FPGA), we catch translocating molecules in the act of crossing between the two nanopores, reversing the bias at one pore so that the molecule is caught in a ``tug-of-war''\cite{chou} between the opposed electrophoretic forces (Fig.~\ref{fig:1}).  Active tug-of-war control suppresses folded translocation as the opposed forces pull the DNA taut between the pores, preventing the propagation of folds between the pores.   Active control enables tuning of the opposing forces to directly control the ``sliding'' velocity at which the molecule contour passes through one pore to the other.  Reducing the sliding velocity increases the tug-of-war duration, the time the molecule spends actively extended with one end in each pore.  We find that the mean tug-of-war duration as a function of reversed bias exhibits a resonance like structure, with the peak lifetime corresponding to a balance of the opposing forces and a physical scenario whereby DNA motion is controlled entirely by diffusional sliding between the pores.  We can implement the full control logic, leading to a controlled tug-of-war, with 76.8\% efficiency on molecules that are initially detected by one pore.
   
   The tug-of-war state is highly unusual:  a non-equilibrium regime arising under low net forcing (low sliding velocity).  We show that the resonance-like enhancement of the tug-of-war lifetime can be modelled using 1D first passage theory.  This result provides a clear physical basis to the degree of lifetime enhancement that is possible using two-pore tug-of-war and quantifies how the enhancement depends on applied voltage at the pores.  Moreover, we find that the diffusion process is in fact sub-diffusive, as has been predicted by translocation simulations,\cite{poresimreview, dubbeldam} giving rise to distributions of tug-of-war lifetime with a stretched exponential character.  The sub-diffusive character of the physics ensures a greater number of long-time tug-of-war states than would be predicted on the basis of simple diffusion.  Lastly, we demonstrate that we can detect correlated motion of bound streptavadin tags at pores 1 and 2 on molecules undergoing two-pore tug-of-war, demonstrating enhanced detection of the tags over a single-pore measurement of the same molecules, and the ability to use physical tags to independently assess translocation velocity.
   
      States where molecules are simultaneously threaded between two pores have been previously demonstrated using a double-barreled glass pipette implementation and closely-spaced nanopores on nitride membranes.\cite{Pud:2016gs, cadinu2017double, cadinu2018double}  Critically, this previous work did not apply an active control logic that could catch the molecule during translocation and reverse the bias.  In consequence, while slow-down did occur, a substantial sliding bias still existed, so that these experiments were far off force-balance, achieving a much smaller slow-down than we demonstrate is possible.  In particular, the average life-time of their tug-of-war state was on order of 10\,ms \cite{cadinu2018double}, sixteen times lower than the average we achieved on resonance, which was 162\,ms, with some of our extreme events lasting longer than a second.

\section{Results:  Two-Pore Tug-of-War}

\begin{figure}[!ht]
\centering
\includegraphics[width=\linewidth]{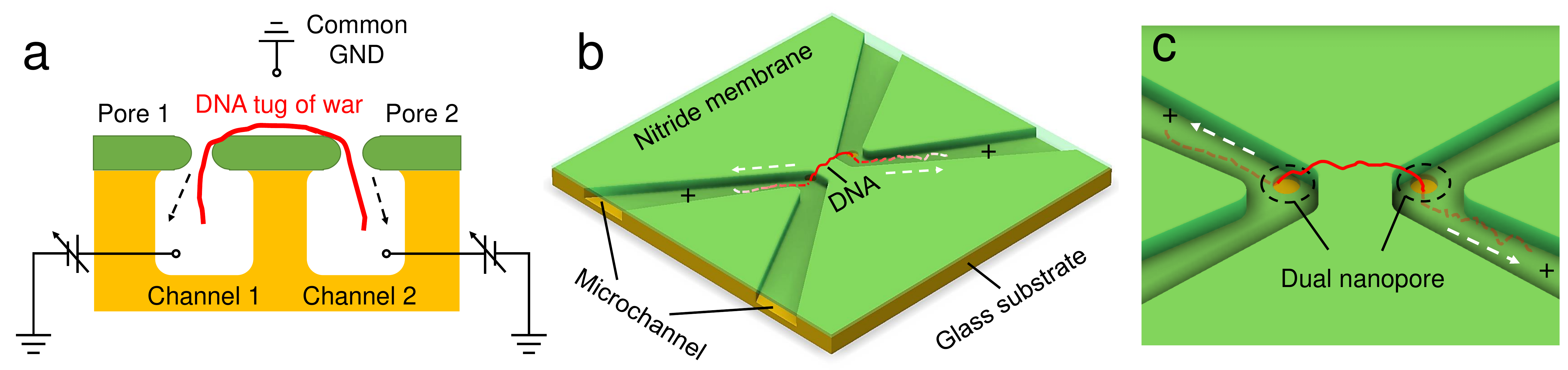}
\caption{\textbf{The dual-nanopore device is comprised of two microfluidic channels sealed by a common membrane containing two closely-spaced nanopores}.  \textbf{a}. Side-view of the two pore device.  A dual-channel amplifier provides independent voltages ($V_1,V_2$) to the reservoirs located adjacent to pore 1 and 2.  A single DNA molecule is captured in the two nanopores and application of competing voltages at pore 1 and 2 results in a ``tug-of-war'' scenario.  \textbf{b}. 3D view of the device. The dual microfluidic channels, with a depth of 1.5\,$\mu$m, are formed in the glass substrate and covered by a 36\,nm SiN membrane. 
\textbf{c}. Zoomed-in 3D view of the dual nanopore device showing the pore locations and a single DNA molecule trapped and extended in a tug-of-war. A detailed fabrication process of the dual-nanopore device is presented in both supplementary materials and elsewheres \cite{Zhang:4Kttepyo, Xu:2017}.} 
\label{fig:1}
\end{figure}

     Each nanopore in the two pore device has a uniquely accessible fluidic channel, and a common fluidic chamber that is accessible to both (Fig.~\ref{fig:1}).  A two channel voltage-clamp amplifier applies voltage across and measures the current through each pore, with the common ground electrode in the common fluidic chamber and source electrodes in each fluidic channel.  Our device architecture minimizes the shared resistance path from each pore to the common ground electrode, and thereby minimizes coupling in the current sensors at constant voltage (Supplemental Sec.~4). Analysis of a representative wafer of the devices showed 73\% yield (61 working out of 83 chips, Supplemental Fig.~S2). Analysis of the two currents at constant voltages also permitted estimation of pore diameters (Supplemental Fig.~S3), and characterization of device noise and capacitance (Supplemental Fig.~S4).

\begin{figure}[!hbt]
    \centering
    \includegraphics[width=\linewidth]{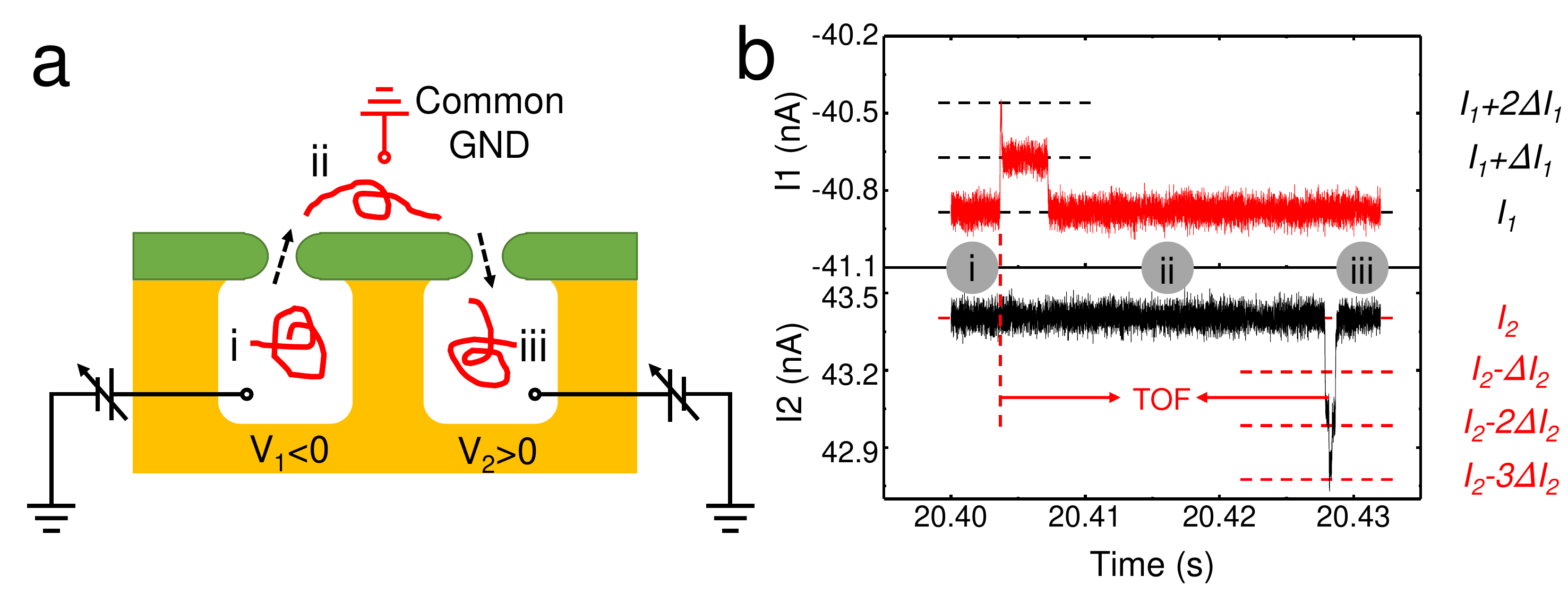} 
    \caption{\textbf{Representative time-of-flight (TOF) event profile with constant voltages (no control)}. \textbf{a}.  Schematic of the path of taken by a single DNA molecule transiting between the pores:  i) in channel 1 prior to capture by pore 1; ii) in the common chamber after capture by pore 1 and prior to capture by pore 2 and; iii) in channel 2 following capture by pore 2.  \textbf{b}.  The trans-pore currents measured at pore 1 and pore 2 ($I_1,I_2$) at applied voltages $V_1=-500$ mV and $V_2=500$ mV.  Labels (i-iii) correspond to the path snapshots shown in (a), enabling determination of the molecule time-of-flight.  The effective diameter of pore 1 is 29.3\,nm and the diameter of pore 2 is 30.4\,nm with an inter-pore distance of 0.62\,$\mu$m.  Time-of-flight statistics are studied in greater detail in a companion manuscript \cite{Zhang:4Kttepyo}.}  
    \label{fig:2}
    \end{figure}

We first present results on use of the device without control in which both voltages remain constant during the experiment. Initially, 48.5\,kbp $\lambda$-DNA molecules are loaded into channel 1 and are captured by pore 1 using a negative $V_1$ polarity. Meanwhile, positive $V_2$ polarity is used to capture DNA from the common chamber by pore 2 and into channel 2. Thus, only the DNA that has already passed through the first pore and into the common chamber is accessible to pore 2. Of course, not all molecules captured through the first pore and into the common chamber are subsequently captured by pore 2. When these do occur, they produce a ``time-of-flight" event pair (Fig.~\ref{fig:2}). The likelihood of a time-of-flight event pair, and the time-of-flight value itself, are primarily functions of the inter-pore distance. Naturally, as the pores come closer together, the likelihood of second pore capture following first pore exit increases. This is described in detail in a companion paper \cite{Zhang:4Kttepyo}.  Inter-pore distances for devices used in this paper are all less than 0.8\,$\mu$m (Supplemental).

The control logic has two functions: first, to promote co-capture of a single DNA into both pores; and second, to promptly engage competing voltage tug-of-war on the co-captured DNA until it exits the pores. The control logic was implemented on a Field Programmable Gate Array (FPGA) for rapid (1\,$\mu$sec) response. In designing the logic, we exploited the fact that the two current signals are decoupled; that is, capture and exit of DNA into pore 1 affects only $I_1$, and the same goes for pore 2 and $I_2$. As is the case for the data in figure \ref{fig:2}, $\lambda$-DNA molecules are loaded into channel 1 and each DNA is initially captured at the first pore using a negative $V_1$ polarity. 

\begin{figure}[!hbt]
    \centering
    \includegraphics[width=\linewidth]{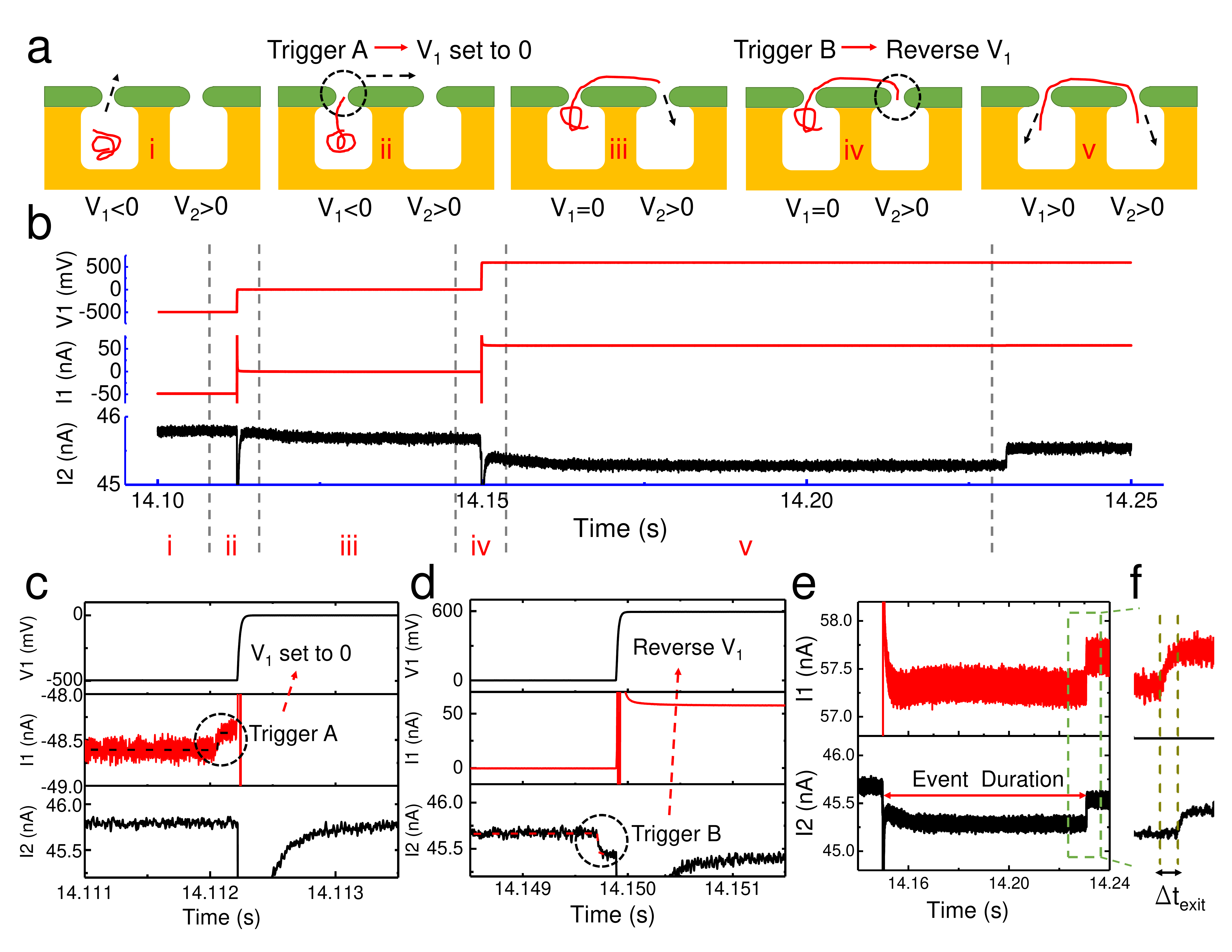} 
    \caption{\textbf{Competing voltage control logic and representative signals during two-pore tug-of-war}. \textbf{a}.  Graphical depiction of the molecule and voltage settings for a subset of the control logic stages. Stage (i) depicts initial capture of a DNA from channel 1 into pore 1.  At 100 $\mu$s after capture of the molecule by pore 1, the FPGA sets $V_1 = 0$ while waiting for pore 2 capture (ii, Trigger A).  Stage (iii) depicts initial capture of a DNA from the common chamber into pore 2.  At 100 $\mu$s after capture of the molecule by pore 2, the FPGA sets $V_1 = V_{1,\text{reverse}} > 0$ (iv, Trigger B), initiating the tug-of-war state (v) which lasts until the molecule disengages from either pore 1 or pore 2.  Failure to reach a stage triggers a reset. \textbf{b}. The trans-pore current signals ($I_1,I_2$) and $V_1$ during a tug-of-war event. \textbf{c}.  Close-up of the detection of molecule capture by pore 1 and subsequent Trigger A (i-ii). \textbf{d}.  Close-up of the detection of molecule capture by pore 2 and Trigger B (iii-iv). \textbf{e}.  Close-up of trans-pore current ($I_1,I_2$) during the competing voltage co-capture event and the co-capture duration (tug-of-war state). After 80\,ms, the molecule disengages from pore 1 and then pore 2, exiting into channel 2.  The device used for this experiment had a pore 1 diameter of 32.7\,nm, a pore 2 diameter of 31.5\,nm and an inter-pore spacing of 0.73\,$\mu$m.}  
    \label{fig:3}
    \end{figure}

Figure~\ref{fig:3}a shows the operation of the control logic. First, capture of a DNA from channel 1 and into pore 1 (Fig.~\ref{fig:3}a i) is detected by the FPGA logic as a drop in baseline of at least 100\,pA for a period of 100\,$\mu$sec, after which time the logic sets $V_1 = 0$ (Fig.~\ref{fig:3}a ii, Trigger A). The rationale for turning $V_1$ ``off'' after 100\,$\mu$sec is to arrest motion of the DNA through pore 1, while making a free end of the DNA sufficiently available in the common chamber for pore 2 to capture.  As with DNA detection at pore 1, capture of a DNA at pore 2 is detected by the FPGA logic when a drop in baseline of at least 100\,pA lasts 100\,$\mu$sec. Following capture of the DNA at pore 2, the FPGA logic sets $V_1 = V_{1,\text{reverse}} > 0$ (Fig.~\ref{fig:3}a iv, Trigger B) to create competing voltages (i.e., tug-of-war, Fig.~\ref{fig:3}a v) that pull the DNA in opposing directions.  Across separate experiments from three different devices, a total of 1,774 out of 2,311 events (76.8\%) captured in pore 1 resulted in co-capture (Supplementary Tables 2, 3 and 4). Failed co-captured occurred when, for example, the DNA captured in pore 2 was no longer in pore 1 (other failure modes are discussed in the Supplementary Material). Once the molecule exits either pore, following a successful co-capture or not, the FPGA resets the control logic for capture of a DNA from channel 1. The $V_2$ voltage is never changed, but we vary the magnitude of the competing $V_{1,\text{reverse}}$ value and assess its influence on the tug-of-war.  Figure~\ref{fig:3}b-f shows the control logic and representative signals during a co-captured tug-of-war event.

\begin{figure}[!hbt]
    \centering
    \includegraphics[width=1\linewidth]{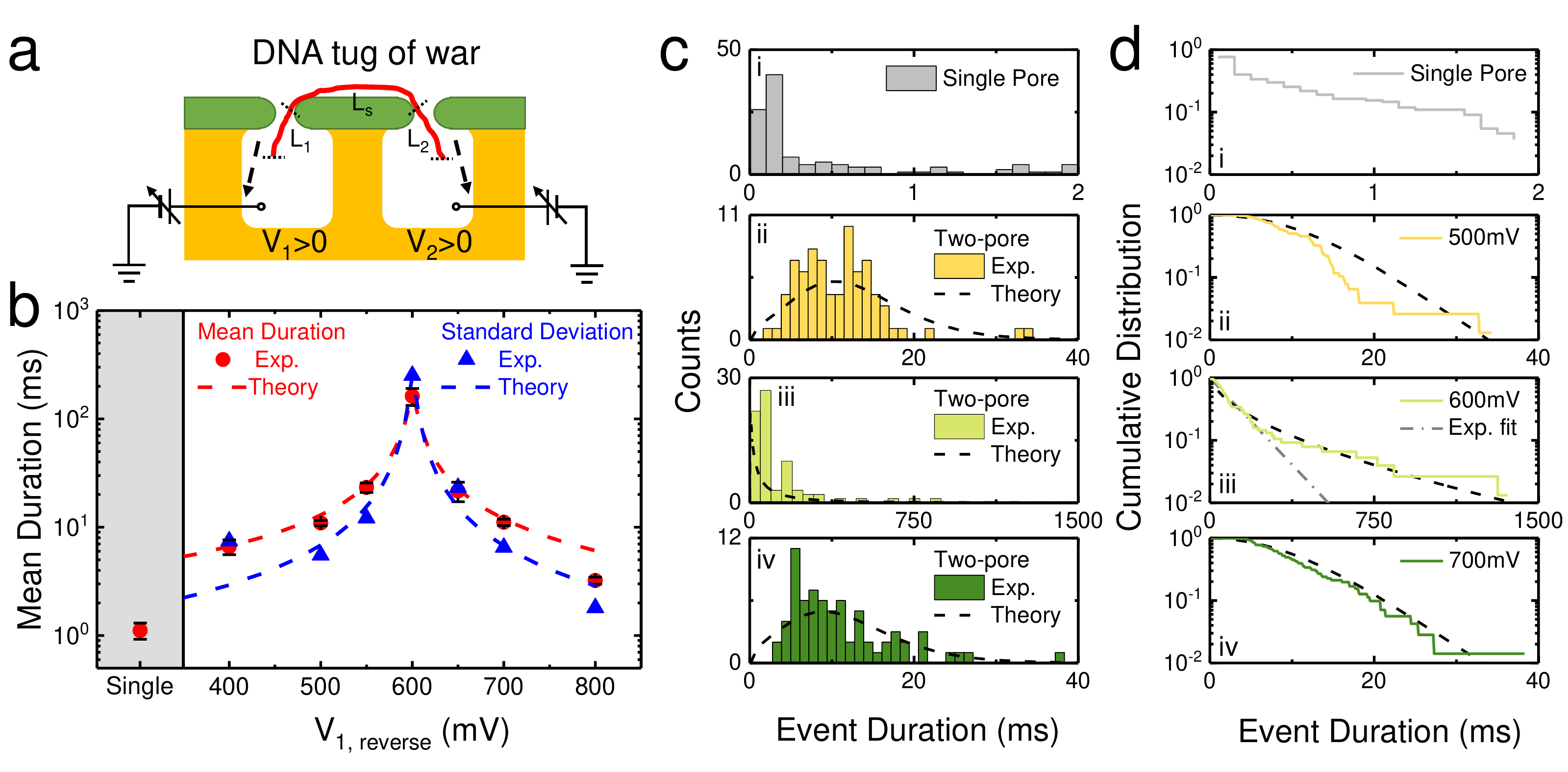} 
    \caption{\textbf{Reducing DNA speed through a nanopore by applying competing voltages}. \textbf{a}.  Schematic of a tug-of-war state, indicating the corse-graining of the DNA contour.   \textbf{b}.  Mean lifetime of tug-of-war state (red) and standard-deviation of tug-of-war lifetimes (blue) as a function of competing $V_1$ voltage, denoted $V_{1,\text{reverse}}$ (same device as in figure 3, with a 32.7\,nm pore 1 diameter, a 31.5\,nm pore 2 diameter and an inter-pore spacing 0.73\,$\mu$m, $V_2 = 500$\,mV).  Error-bars on mean-value measurements correspond to standard error on the mean.  \textbf{c}. Distributions of tug-of-war lifetimes for single pore (i), $V_{1,\text{reverse}}=500$\,mV (ii), $V_{1,\text{reverse}}=600$\,mV  (iii, peak duration) and $V_{1,\text{reverse}}=700$\,mV (iv).  \textbf{d}.  Cumulative distributions of tug-of-war lifetimes for single pore (i), $V_{1,\text{reverse}}=500$\,mV (ii), $V_{1,\text{reverse}}=600$\,mV (iii, peak duration) and $V_{1,\text{reverse}}=700$\,mV (iv).  Note that the cumulative distributions are defined by integrating all events from duration $t$ to duration $t=\infty$; the distributions are normalized to unity at $t=0$.  Fit to theoretical predictions in (b), (c) ii-iv and (d) ii-iv shown as dashed black line.}
    \label{fig:4}
    \end{figure}

Applying competing voltages to the pores reduces the DNA speed and increases the duration of the tug-of-war (Fig.~\ref{fig:4}).  The average tug-of-war lifetime is a function of competing voltage $V_{1,\text{reverse}}$ (Fig.~\ref{fig:4}b, for comparison we also show the average single-pore translocation time through pore 1 on the same device).  Figure.~\ref{fig:4}c-d shows the distribution of event life-times, both as binned event duration (Fig.~\ref{fig:4}c) and cumulative binned event duration (Fig.~\ref{fig:4}d).  The pore 2 voltage $V_2 = 500$\,mV was kept constant allowing a minimal perturbation of $I_2$ during application of the control logic.   Remarkably, we find that the life-time of the tug-of-war state is maximized at $V_{1,\text{reverse}} = 600$\,mV with an average duration $\sim$160  times longer than the single pore duration.  Qualitatively, this effect arises as the sliding velocity of the DNA from one pore to the other is canceled when equal forces are applied to the DNA at ether pore, maximizing the tug-of-war duration.   Far off force-balance, the DNA will slide quickly from pore 1 to pore 2 (for $V_{1,\text{reverse}} < 600$\,mV), or from pore 2 back through pore 1 (for $V_{1,\text{reverse}} > 600$\,mV), leading to low tug-of-war lifetimes.   In two additional experiments performed with different devices, the duration maximizing values were $V_{1,\text{reverse}}=520$\,mV and 550\,mV.  It is not surprising that the competing voltages at peak lifetime were not equal, since the pair of pores in each experiment differed in size and shape and thus distributed the field force differently.\cite{Pud:2016gs} The average increase in average duration from single pore to competing two pore was 114-fold across the three experiments.  We also observed a small number of very long events (1000X longer than the average single pore translocation time or $>$1 sec, Supplemental Fig.~S10).  While these long events were relatively rare, their frequency of occurrence was comparable to the total fraction of random two-pore co-capture events in Pud \emph{et al}.\cite{Pud:2016gs}  
In addition, the decoupled current measurements do permit accurate tracking of the direction and time scales of exit, and as expected the DNA tends to exit in the direction of higher force (Supplemental Fig.~S6j. Fig.~S8c and Fig.~S9c).

  When $\lambda$-size DNA passes through a large ($>5$\,nm) nanopore, the probability of folding occurring during translocation is high ($\sim$60-70\% for pores in the 10-14\,nm range.\cite{dekkerfold,keyserbarcode1, keyserbarcode2}).  We find active tug-of-war control significantly reduces the folding probability, even for the large ($>25$\,nm) pores used here.  Under optimum conditions more than 79\% of events with active tug-of-war are unfolded and almost all the remaining folded events decay to unfolded events while tug-of-war control is maintained.
  
\begin{figure}[!hbt]
    \centering
    \includegraphics[width=0.9\linewidth]{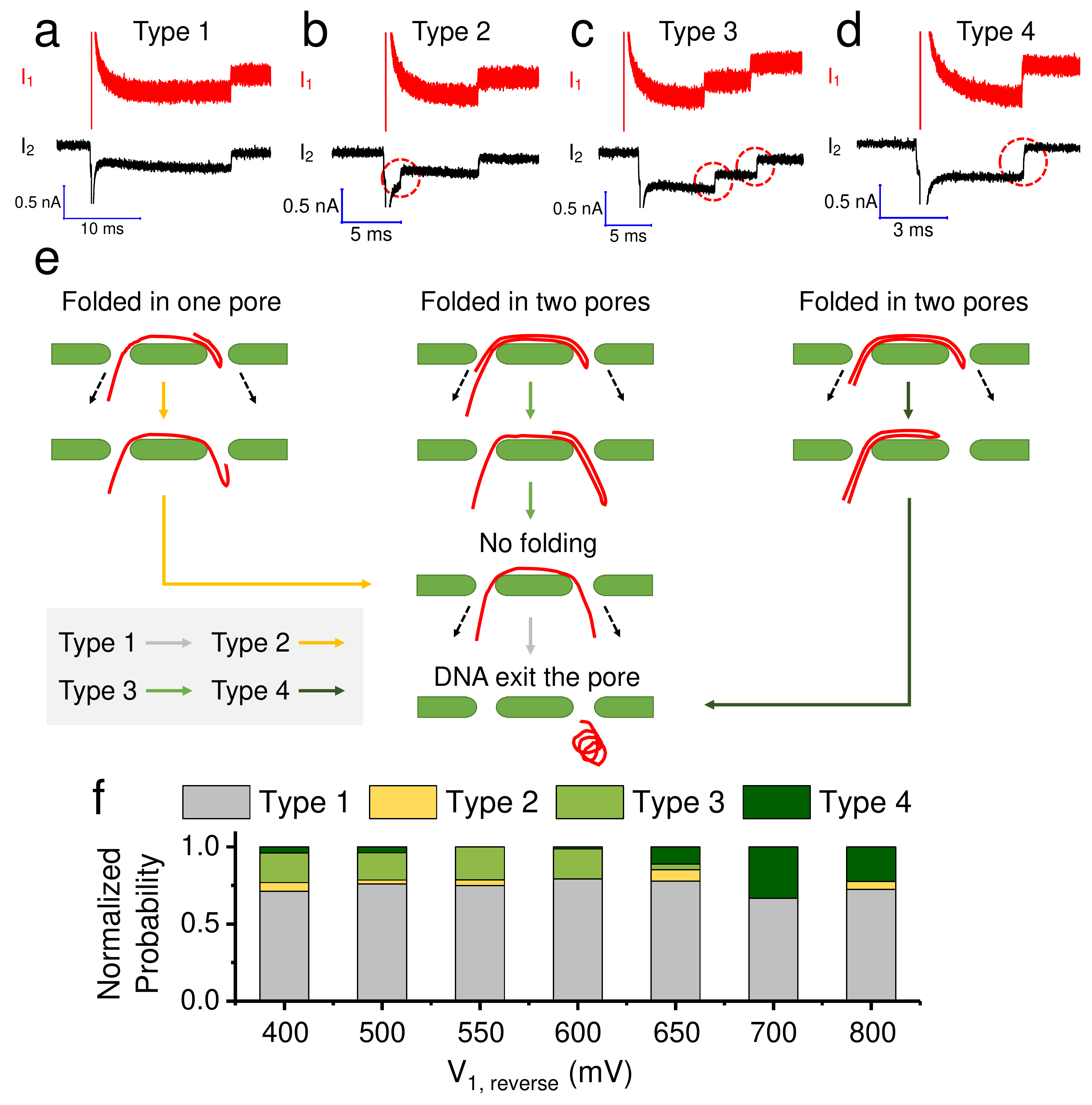} 
    \caption{\textbf{Molecular folding and unfolding under two-pore control}. \textbf{a-d}. The four types of double-channel current profiles shown correspond to different molecular folding events in the two pore experiment: (a) Type 1 -- no folding; (b) Type 2 -- folded in one channel (always channel 2) and decays to unfolded tug-of-war; (c) Type 3 -- folded in two channels and decays to unfolded tug-of-war; (d) Type 4 -- folded in two channels and decay terminates tug-of-war.  \textbf{e}. A cartoon depicting possible decay paths under two-pore control.   \textbf{f}. Normalized probability distribution of different event types as a function of reverse bias $V_\text{1, reverse}$.   Folding data is collected from the same device as in figure 3 and 4.}
    \label{foldfig}
    \end{figure}

       In contrast to a single pore device,  co-captured molecules in a two-pore device can be folded at both pores, only one pore, or neither pore.  Events which are folded at neither pore correspond to the example shown in Fig.~\ref{fig:3}b-f and Fig.~\ref{foldfig}a (we designate these events as Type 1).  Figure~\ref{foldfig}b shows an example of an event folded at only one pore and Fig.~\ref{foldfig}c-d shows examples of events folded at both pores.  When folding occurs at only one pore (Type 2), than the folded tug-of-war state will quickly decay to an unfolded tug-of-war state (a molecule folded at only one pore will possess a free strand, lying between the pores, that experiences an unbalanced force rapidly leading to its unraveling, see Fig.~\ref{foldfig}e).  If the folding occurs at both pores, then the most likely scenario is that a single fold exists on the chain that simultaneously extends through both pores (Fig.~\ref{foldfig}e).  While a conformation possessing two separate folds, each fold passing through a separate pore, is possible, the free ends of the molecule in such a state would lie in between the pores and lead to a rapid unraveling of the folds (just as for a Type 2 event).  For a conformation possessing a single fold that extends through both pores two possible decay scenarios are possible.  Firstly, decay can occur through one of two the molecule free ends passing through a pore, resulting in an unfolded tug-of-war state (Fig.~\ref{foldfig}e, Type 3).  Secondly, decay can occur through the molecule hooked end passing through a pore, leading to termination of the tug-of-war (Fig.~\ref{foldfig}e, Type 4).   Figure~\ref{foldfig}f shows the distribution of different event types as a function of reverse bias.  Note that the number of events of Type 2 is always quite low, reflecting a high degree of correlation in the conformational behaviour between the pores.  The number of events of Type 3 is high for $V_{1,\text{reverse}}$ below the bias corresponding to the peak lifetime.  The number of events of Type 4 is high for  $V_{1,\text{reverse}}$ above the bias corresponding to the peak lifetime.  This behaviour reflects the well-known phenomena that double folds form preferentially at the molecule leading edge during pore translocation.\cite{dekkerfold}  A fold formed at pore 1 will then propagate through to pore 2, so that the free ends are located before pore 1, and the hooked end is present on the far side of pore 2 (as represented in Fig.~\ref{foldfig}e for the folding state with folds in both channels).  When $V_{1,\text{reverse}}$ is set below the bias corresponding to peak lifetime, this will pull DNA from pore 1 to 2, resulting in one of the free edges tending to escape, so that the tug-of-war will be maintained (leading to more events of Type 3).  When $V_{1,\text{reverse}}$ is set above the bias corresponding to peak lifetime, this will pull DNA from pore 2 to 1, resulting in the hooked end escaping, so that the tug-of-war will be terminated leading to more events of Type 4).  The total number of events of Type 3 and Type 4 is roughly conserved across the biasing range, suggesting that the summed Type 3 and 4 events reflects the total number of events with folds at both pores that are created during steps i-iv initiating the tug-of-war.  We find that the number of events of Type 3 and Type 4 is sensitive to pore spacing, with devices with lower pore spacing having a still high but lower number of unfolded (Type 1) translocations (for device 2, with a spacing of 675\,nm, a maximum 73\% are unfolded and for device 3, with spacing of 624\,nm, a maximum 55\% are unfolded, compared to a maximum of 79\% of events unfolded for device 1 with an inter-pore spacing of 734\,nm).

\section{Model:  First Passage Time for Tug-of-War}

      Here, using a first passage theory adopted from approaches developed to study single-pore translocation, we demonstrate quantitatively how reducing DNA sliding velocity in the vicinity of force-balance leads to an enhancement of the tug-of-war lifetime.  We first course-grain the molecular configuration in a tug-of-war state into three bins:  the contour in reservoir 1 $L_1$, the contour extended between the pores lying above the membrane $L_s$ and the contour in reservoir 2 $L_2$ (Fig.~\ref{fig:4}a).  With the constraint imposed by the molecule's fixed contour length $L=L_1+L_2+L_s$, there are two independent variables.  The voltage applied at each pore gives rise to electrophoretic forces on the DNA, which are balanced by frictional forces and the DNA chain's elastic response. 
      
    We choose to focus our theory on the tug-of-war (Fig.~\ref{fig:3}a v), with the earlier control steps providing initial conditions for the tug-of-war state that will be treated as adjustable fitting parameters (e.g. the distribution of DNA between the bins $L_1$, $L_2$ and $L_s$ upon application of competing voltages).  Given an $\sim 1$\,ms average single pore translocation time at 500\,mV, we estimate that only a small portion of the molecule ($\sim$10\%) can cross pore 1 during the 100\,$\mu$s long trigger A and enter the space between the pores.  Yet, in the time after the pore 1 voltage is zeroed but prior to the molecule entering pore 2 (Fig.~\ref{fig:3}a iii), additional molecule contour will accumulate around pore 2.  This accumulation of contour arises due to the net force applied to the molecule by the electric field in the vicinity of pore 2.  The amount of contour transferred across pore 1 during control step iii is unknown, but it could be substantial due to the length of step iii ($\sim10$'s of ms).
    
    During application of trigger B, another $\sim$10\% of contour will cross pore 2.  At this point, competing voltages conditions are applied.  A transient process will occur, on order of the single pore translocation time, whereby contour is transferred from $L_s$ to $L_1$ and $L_2$; this process will terminate when the elastic restoring force exerted by the extended DNA balances the applied electrical forces.  Critically, the electrical forces are sufficiently strong to almost completely extend the DNA between the pores.  For example, measurements of nanopore induced electrophoretic force using optical tweezers\cite{vandorpNatPhys} show that pores of the size used exert a mechanical force of around 0.13\,pN/mV, giving rise to a force of over 10\,pN for even the smallest voltages applied.  This force vastly exceeds the force threshold $f^{\star}=k_B T/P\approx0.08$\,pN required for high chain extension (with $P=50$\,nm the DNA persistence length).\cite{Marcosiggia}  Consequently, we argue that contour will be pulled from $L_s$ into $L_1$ and $L_2$ until the segment $L_s$ is stretched tight between the pores.  At this point, fluctuations in $L_s$ are suppressed and $L_s$ is approximately equal to the pore spacing $d$ so that $L_s$ can be treated as a fixed parameter, reducing the problem to one dimension.  The effect of the control conditions are thus to provide a single initial condition for the tug-of-war state:  the ratio of contour distributed between the pore reservoirs.  The extended chain, pulled taut between the pores, will experience a sliding velocity $v$ proportional to the difference in electrophoretic forces induced by the pores ($f_1$ and $f_2$).  The difference in electrophoretic forces is in turn proportional to the voltage drop at the pores: $v=\left(f_2-f_1 \right)/\xi$ with $\xi$ a friction factor.   Letting $f_i=a_i V_i $, with $a_i$ a parameter that relates voltage drop to applied force at pores 1 and 2, we have $v=A \left(V_2-V_{\star} \right)$.  The parameter $A \equiv a_2/\xi$ relates voltage to velocity scale and $V_{\star} \equiv (a_1/a_2) V_1$ determines the voltage corresponding to the peak lifetime of the tug-of-war lifetime curve (Fig.~\ref{fig:4}b).
        
     First passage approaches have a long history in modelling the time-scale for chain translocation through a single-pore.\cite{poresimreview, porereview2}  In the classic model introduced by Lubensky \emph{et al},\cite{lubensky, Li2010, Ling2013} the translocation process is viewed as a 1D biased random walk.   The probability $P(s, t)$ for contour $s$ to be found on the trans-side of the membrane at time $t$ is determined by solving a Fokker-Plank equation\cite{Ling2013, polson} 
\begin{equation}
\frac{\partial P}{\partial t}=D \frac{\partial^2 P}{\partial s^2}-v \frac{\partial P}{\partial s} \label{eq:FP1}
\end{equation}   
with $D$ a diffusion constant and $v$ a convective velocity determined by the electric field at the pore.   Equation~\ref{eq:FP1} is solved subject to an absorbing boundary condition $P(L,t)=0$, reflecting DNA chain escape upon complete translocation.  The distribution of first passage times is determined from: $\mbox{FP}(t)=-(d/dt)\int_0^L P(s, t)\,ds$.   This approach has been successfully used to model experimental distributions of DNA translocation times through single solid-sate pores.\cite{Li2010, Ling2013}.

  The single-pore translocation model can be mapped onto our tug-of-war scenario, letting $s \equiv L_2$, which follows from our argument that contour $L_s$ is stretched tight and fixed.  We use the initial condition $P(s, 0)=\delta (s-s_o)$ with $s_o \equiv L_{o2}$ fixing the amount of contour in reservoir 2 at $t=0$ ($L_{o2}$ is determined experimentally by the amount of contour in reservoir 2 when trigger B is initiated).   Absorbing boundary conditions $P(L-L_s, t)=0$ and $P(0,t)=0$ force the tug-of-war state to terminate when the molecule escapes from either pore 1 ($s=L-L_s$) or pore 2 ($s=0$) respectively.   These conditions together specify a convective-diffusion model on a finite interval that can be solved analytically\cite{farkas2001one}, leading to a first passage time that is maximized in the absence of convection ($v=0$) and a mean first passage time for $v=0$ of the form:  $\tau_{\mbox{\scriptsize max}}=B (B-1) L_o^2/D$ with $B \equiv L_{o2}/L_o$ and $L_o \equiv L-L_s$.
    
  Simple biased diffusion, however, leads to a quantitatively incorrect prediction for the lifetime of the measured tug-of-war states.  In particular, simple diffusion predicts that the distribution of first passage times at $v=0$ should have an exponential tail in the limit of large first passage times, yet our data has a stretched exponential character with a large number of long-time events (Fig.~\ref{fig:4}d iii).   Consistent with this observation, we find that the measured standard-deviation is greater than the distribution average.   This suggests that the tug-of-war process is in fact \emph{sub-diffusive}, as has been predicted by simulations of force-free translocation,\cite{poresimreview, dubbeldam} due to the influence of coupled polymer modes on the fluctuations of the translocation velocity at the pore.\cite{anomdiff,dubbeldam}
  
       To include the sub-diffusive character of the motion in our first passage theory, we use an approach developed by J. Dubbeldam \emph{et al}, who derived a Fokker-Plank equation describing sub-diffusive translocation physics within the framework of fractional Brownian motion.\cite{dubbeldam}  In this approach, the tug-of-war dynamics is governed by the Langevin equation
     \begin{equation}
     \frac{ds}{dt}=v(t)
     \label{eq:FBM1}
     \end{equation}
with $v(t)$ a fluctuating translocation velocity at pore 1.  Simulations suggest that $v(t)$ obeys Gaussian statistics but that $s(t)$ can have a variance that is sub-diffusive, e.g. $\langle \left(s-s_o \right)^2 \rangle \sim t^{\alpha}$ with $\alpha<1$.\cite{poresimreview, poresim1}  J. Dubbeldam \emph{et al} show that if $v(t)$ has a fixed average $v=\langle v(t) \rangle $, Eq.~\ref{eq:FBM1} leads to a Fokker-Plank equation of the same structure as Eq.~\ref{eq:FP1} but possessing a \emph{time-dependent} diffusion constant $D(t)=\int_0^t v(t^{\prime})v(0)\,dt^{\prime}$ fixed by the correlation function of the fluctuating translocation velocity.  The ansatz $D(t) \equiv D_o t^{\delta}$, with $D_o$ a constant, is equivalent to $\langle \left(s-s_o \right)^2 \rangle \sim t^{\alpha}$ with $\delta=\alpha-1$ (note $\alpha=1$ and $\delta=0$ correspond to the case of simple diffusion).  In the limit that $v=0$, this approach explicitly leads to a stretched exponential first passage time distribution, $\mbox{FP}(t) \sim t^{\alpha-1} \exp \left [ -(\pi^2/L^2) D_o t^{\alpha}  \right]$.  
  
  A combination of simulation and scaling arguments determine the exponent values.    In the long-time limit, the DNA chain slides over its entire contour length on times on order of the largest relaxation time of the chain.  For a self-avoiding polymer in the absence of hydrodynamic interactions this yields $\alpha=2/(2 \nu+1)\approx 0.92$ ($\nu=0.59$ is the exponent characterizing a self-avoiding polymer in 3D).\cite{scaling}  At times below the largest chain relaxation time, all polymer modes contribute, yielding $\alpha=0.55$ if hydrodynamic interactions are absent.\cite{dubbeldam,poresimreview}  Brownian-dynamics simulation explicitly shows that $\alpha$ undergoes a transition between these limits as simulation time increases.\cite{dubbeldam}  Thus, we expect $\alpha$ to lie between 0.55 and 0.92.
  
  To build sub-diffusion into our model, we solve Eq.~\ref{eq:FP1} subject to the boundary conditions previously described but with a time-dependent diffusion constant.  For specificity, we set $D(t) \equiv D_o (t/\tau)^{\delta}$; the constant $D_o$ has dimensions of diffusion and $\tau \equiv L_o^2/D_o$ is a diffusive time-scale.  An analytic solution does not exist for $\delta<0$ if the convective term is present, thus we solve Eq.~\ref{eq:FP1} numerically (using the commercial package FlexPDE).  First passage solutions are obtained as a function of parameters $v$, $B$ and $\delta$ in units of rescaled time $T \equiv t/\tau$ and contour $S=s/L_o$.   Interpolation performed in Matlab is then used to construct solutions in physical units as a function of $A$, $B$, $D_o$, $\delta$ and $V_{\star}$.  Using nonlinear least squares, we fit our model with equal weighting to the mean and standard-deviation of the lifetime distribution over the range of voltages used (see Fig.~\ref{fig:4}b, fit is dashed black line).  The model accurately predict the experimental data, with resulting best-fitting parameter values:  $A=6\pm1$\,$\mu$m s$^{-1}$mV$^{-1}$; $B=0.5\pm0.1$; $D_o=20\pm14$\,$\mu$m$^2$/s; $\delta=-0.43\pm0.06$ and $V_{\star}=600\pm6$\,mV. 
  
   While our model contains five fitting parameters, these parameters also describe the full distributions of tug-of-war lifetimes (see Fig.~\ref{fig:4}c-d).  These fitting parameters also have a clear qualitative interpretation:  (1)  $V_{\star}$ determines the peak voltage; (2) $B$ determines the degree of asymmetry between the two sides of the lifetime curve (Fig.~\ref{fig:4}c); (3) $\delta$ determines the degree of sub-diffusion, leading to the distribution ``stretching'' and departure from exponential behaviour; (4) $D_o$ determines the maximum lifetime and (5) $A$ determines the conversion between voltage difference and sliding velocity, controlling the ``narrowness'' of the lifetime curve on the experimental voltage scale.  The error-bars of $D_o$ and $\delta$ are large, because these parameters are strongly correlated in the fit, but the results are consistent with sub-diffusive behaviour and yield an $\alpha=\delta+1=0.57\pm0.06$, suggesting we are closer to the low-time limit.  This is expected as the DNA chain is partitioned between the two reservoirs.  Full translocation of the chain portion through one of the pores will involve a fraction of the chain contour and can thus occur on times-scales below the largest chain relaxation time (the time-scale for sliding of the \emph{entire} chain).  The value of $D_o$ is large, but gives diffusivities on order of those measured for single-pore translocation \emph{et al};\cite{Ling2013, BLu2011} these large values of diffusivity arise as only a small portion of the contour very close to the pore participates in the non-equilibrium motion (i.e. corresponding to a tension blob in the pore vicinity).
  
\begin{figure}[!hbt]
    \centering
    \includegraphics[width=0.7\linewidth]{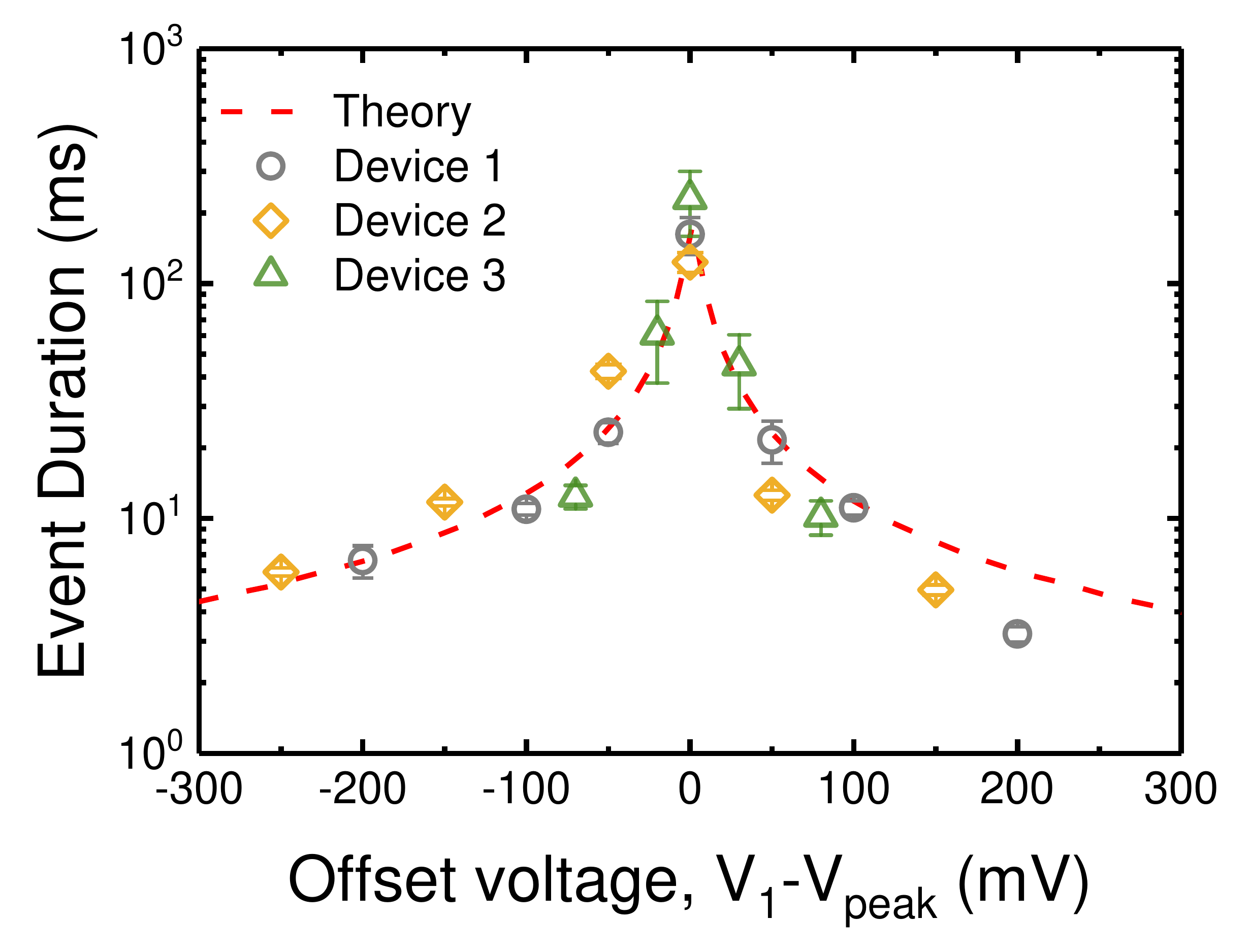} 
\caption{\textbf{Average tug-of-war duration for all two-pore devices versus offset voltage}.   We rescaled all experimental data (3 data sets in total, each taken for a different device) by plotting event duration against offset voltage (V$_{1}$-V$_\text{peak}$).  This rescaling effectively collapses the data sets.  Our theoretical model agrees well with the rescaled three two-pore experiments, with the exception of points for positive voltage offset greater than 100\,mV, which fall systematically below the model. For device 1, pore 1 has an effective diameter of 32.7\,nm, pore 2 has an effective diameter of 31.5\,nm and the inter-pore distance is 0.73\,$\mu$m (same device used in Fig.~3, 4 and 5).  Device 2 has an effective pore 1 diameter of 27.8\,nm, a pore 2 diameter of 27.8\,nm and an inter-pore distance of 0.68\,$\mu$m.  Device 3 has an effective pore 1 diameter of 26.2\,nm, a pore 2 diameter of  28.0\,nm, and an inter-pore distance of 0.62\,$\mu$m.  
Effective pore diameter is calculated based on open pore current. Inter-pore distance is obtained from SEM measurement.}
    \label{fig:4addition}
    \end{figure}

      Our fitted theory also predicts the outcome of experiments in the two additional devices tested.   Figure~\ref{fig:4addition} shows the compiled tug-of-war duration for all events versus offset voltage, defined by subtracting the voltage of peak lifetime from each curve.  The agreement between the three rescaled experiments and the model prediction suggests that the main effect of the different device geometry is to shift the peak voltage.  Our model systematically overestimates the experimental lifetimes for large positive voltage offset ($>100$\,mV); however, we believe this disagreement arises because our assumption of using a fixed contour asymmetry ($B$) for all the reverse voltages applied is too crude.  In particular, our model does not take into account the details of the transient process before the contour is pulled taut, a process that also depends on the nonuniform distribution of contour between the pores introduced in control step iii.  
      
      Let us consider how these details might play out in practice.  Note that for negative voltage offset, the bias pulls contour towards pore 2.  As we expect the contour to be concentrated near pore 2 at the end of control step iii, this bias results in most of the DNA between the pores crossing pore 2, and thus a substantial increase of $L_2$ above the 10\% baseline induced by trigger B.  The increasing value of $L_2$, arising for negative voltage offset, leads to the more equal distribution of contour between the pores observed in our fitted value of $B$.  Furthermore, we argue that the tendency is for most of the contour initially present between the pores to be transferred to $L_2$, until a sufficiently large bias towards pore 1 exists at high positive voltage offset.  This large bias towards pore 1 will overcome the spatial asymmetry induced in control step iii and pull more contour to $L_1$ at the beginning of the tug-of-war.  The net result is a lower $B=L_{o2}/L_{o}$ value at high positive voltage offset, which in turn leads to a lower tug-of-war lifetime.
       
\section{Results:  Tug-of-War with Bound Probes}
 
\begin{figure}
    \centering
    \includegraphics[width=0.85\linewidth]{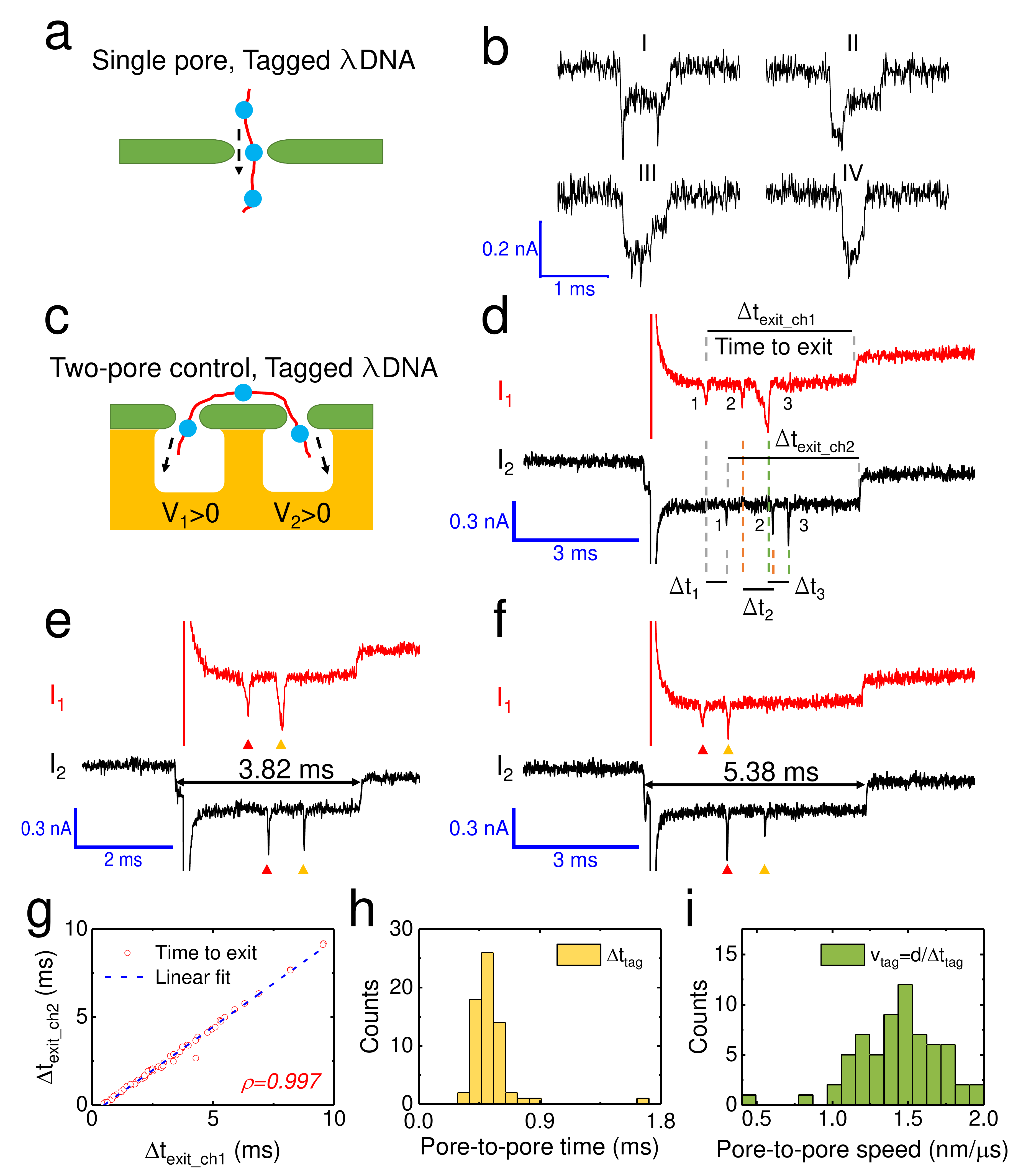} 
    \caption{\textbf{Detection of sequence-specific tags on co-captured and controlled $\lambda$-DNA molecules}. \textbf{a}.  Cartoon of mono-streptavidin (MS) ``tags'' attached to $\lambda$-DNA during a competing voltage co-capture event. Each captured DNA had at most 7 MS tags attached at specific sites along the molecule (Supplemental Figs.~S11 and S12), with most DNA have 3 MS or less according to dual-pore analysis. \textbf{b}. A single pore event with the MS-labeled $\lambda$-DNA still produces fast events (1 ms, -500\,mV), with the tag-induced spikes generally unresolvable. \textbf{c}.  Cartoon of mono-streptavidin (MS) ``tags'' attached to $\lambda$-DNA during a competing voltage co-capture event.  \textbf{d,e,f}  Three representative co-capture ($I_1,I_2$) event pairs with $V_{1,\text{reverse}} = 400$\,mV.  Additional representative events are shown in Supplemental Figs.~S14-S16.  No $\lambda$-DNA co-capture events produced such spikes in the absence of MS.  \textbf{g} Label exit times for channel 2 plotted versus exit times for channel 1.  \textbf{h}.  Histogrammed pore-to-pore transit time; \textbf{i} histogrammed label speeds $d/\Delta t_{\mbox{\scriptsize tag}}$.   The pores used had diameters of 24\,nm and 31\,nm respectively. The pore-to-pore distance is 0.72\,$\mu$m.}
    \label{fig:5}
    \end{figure}

    We can detect mono-streptavidin (MS) tags along translocating $\lambda$-DNA using the two-pore device.  Mono-streptavidin creates a physical ``bump'' on the DNA that leads to a local increase in the blockade signal.   MS-tags for $\lambda$-DNA  undergoing single pore translocation cannot be reliably detected (Fig.~\ref{fig:5}a-b); however, the tags can be clearly observed for a dual-captured molecule under tug-of-war control (Fig.~\ref{fig:5}d-f), demonstrating that the DNA slow-down promotes reliable nanopore sensing of small translocating features, even for the large 20-30\,nm pore diameters used here.  In particular, note folded translocation events prevent reliable detection of tags.   As two-pore control eliminates folding in the majority of translocation events, the technology potentially enables access to a  larger portion of the data.  For comparison, in conventional single-pore electronic barcoding, performed with a 14\,nm pore,\cite{keyserbarcode1, keyserbarcode2} the majority of events are folded and do not possess usable barcodes so that only 30\% of the events can be used.  For the data shown in Fig.~\ref{fig:5}d-f, 331/368 or 90\% of events are unfolded.
    
    The two-pore device can also perform label re-sensing.   We determine the time interval between the time of label detection and the time at which the molecule end exists the pore for channel 1 and 2 ($\Delta t_{\mbox{\scriptsize exit-ch1}}$ and $\Delta t_{\mbox{\scriptsize exit-ch2}}$, Fig.~\ref{fig:5}d).  While the detailed translocation signal induced by a MS-tag can vary between the pores (Fig.~\ref{fig:5}d-f, Supplemental Figs.~S14-S16), the tag positions on the molecule blockade signals for pore 1 and 2 are highly correlated, as revealed by a plot of $\Delta t_{\mbox{\scriptsize exit-ch1}}$ versus $\Delta t_{\mbox{\scriptsize exit-ch2}}$ (Fig.~\ref{fig:5}g).  Moreover, the pore-to-pore travel time for resolvable adjacent pairs of spikes in $I_1$ and $I_2$ ($>5\sigma$ below DNA amplitude) can be used to measure the pore-to-pore transit speed for each tag ($\Delta t_{\mbox{\scriptsize tag}}$, Fig.~\ref{fig:5}d,h), from which the sliding speed of the tag can be determined ($d/\Delta t_{\mbox{\scriptsize tag}}$, Fig.~\ref{fig:5}i).  The average measured translocation speed, $1.43\pm0.03$\,nm/$\mu$s, is in line with the translocation speed we would estimate from our first passage model.  In particular, using the fitted value for $A$, and assuming that the $V_{1,\text{reverse}} = 400$\,mV is 200\,mV from the life-time peak as in Fig.~\ref{fig:4}b, our fitted value of $A$ leads to the estimate $v=1.2\pm0.2$\,nm/$\mu$s.  While the reagents produced DNA that could have a range of MS tags bound, with a maximum of seven per DNA, we typically see no more than 2-3 tags for a given event.  Our first passage model predicts that the contour is initially equally divided between the reservoirs (the fitted value $B=0.5\pm0.1$).  Thus, only half of the molecule barcode will be observed for a given translocation (note that there are four separate clusters of tag sites fairly uniformly distributed over the $\lambda$-DNA, Fig.~S11).  This feature can be addressed via a number of approaches.  Firstly, by increasing the blockade duration required for trigger B to be applied, the control system can be tuned to produce more asymmetric tug-of-war scenarios, with a larger portion of molecule contour on one side of the pore.  Secondly, we can in principle produce a full-map by oscillating the sliding voltage polarity so the molecule rocks back and forth, increasing the duration of each pulse of fixed polarity so a greater portion of the molecule will translocate on each pulse.  Eventually, the molecule will slide too far and disengage completely from one pore terminating the tug-of-war state, but this approach will produce as close to a full map as desired by reducing the pulse time increments.   Finally, we can add molecule end-adaptors featuring large physical tags that will signal the presence of the molecule end.  Detection of these end-tags by either channel will send a trigger signal to reverse molecule sliding direction, maintaining the tug-of-war.

\newpage

\section{Discussion and Conclusions}

     Current solid-state nanopore technology lacks an effective means for combined conformational and dynamic control of translocating single molecules.  Two-pore tug-of-war implemented with active feedback gives rise to predominantly linearized translocation signatures, obviating the folding that has been a significant challenge in electrical barcoding approaches.  While very small (sub 5\,nm) pores do inhibit folding,\cite{mellersmallpore} such small pores are difficult to make in a repeatable and scalable way.  Moreover, two-pore tug-of-war provides a versatile speed control method that can reduce the average sliding velocity of the chain to zero.  Another advantage is the ability to independently sense a given region of the molecule twice with two pores, enabling (1) more accurate sensing via two-channel correlation and (2) independent measurement of DNA sliding velocity, which we demonstrate with MS-tags.   Finally, our control technique enables access to highly novel DNA translocation regimes, in particular the ability to observe low-force translocation scenarios were diffusion dominates.   
     
     The ability to observe diffusion dominated translocation regimes raises an important question for the nanopore field:  what is the \emph{smallest} average translocation speed required to optimize mapping?  Classically, translocation slow-down is desired due to the need to reduce the bandwidth required to resolve the passage of small closely-spaced features.  Yet, if the translocation speed becomes too small, then diffusion will dominate the dynamics of a tug-of-war state,\cite{BLu2011} leading to large random fluctuations of the molecule contour back and forth between the pores.  These random fluctuations will induce significant variation in label separations measured between separate translocation events, even for molecules possessing identical labeling chemistry, making impossible precise mapping of label positions across the ensemble of measured events.  Note that this trade-off is not unique to our system, but exists in any mapping technology where the molecule is not transiently bound (i.e. repeatedly held immobile and released at the molecular scale during the translocation).  In protein pore sequencing technology, such a molecular ratcheting technology is borrowed from biology.  Yet, a molecular ratcheting technology does not yet exist in the context of solid-state pores.  Our control technology enables us to pick a translocation velocity slightly off the life-time peak, enabling us to choose an operating point that is the best compromise between minimizing bandwidth and preventing diffusive broadening.  
       
    Here we apply a \emph{fixed} reverse bias, yet our FPGA logic could potentially implement a \emph{dynamic} reverse bias that is varied during the course of a translocation event (an FPGA controlled amplifier has a response time of $\sim$20\,$\mu$s\cite{Dunbar:2015vn}).  In particular, measurement of the chain sliding velocity during tug-of-war could be used as a control input to dynamically adjust the FPGA-controlled bias to counteract random fluctuations of the chain in and out of the pore (i.e. diffusion).  The ultimate objective of such dynamic control would be to pause translocation at a given sequence position; such \emph{anti-Brownian} control techniques have already been developed to construct traps for small colloidal particles.\cite{cohen}  The dynamic diffusion constant used in the sub-diffusive modelling suggests how such control might be applied in practice:  the dynamic reverse bias would be used to induce anti-correlations in the velocity fluctuations at the pores, thereby reducing the value of the time-integral and driving $D(t)$ to zero.  The key challenge is not performing sufficiently fast control, which is already feasible with our FPGA system, but obtaining densely sampled measurements of chain velocity.   One potential solution to this problem, demonstrated here at a preliminary level, is to use physical labels to perform velocity measurements.
    
    Our two-pore control technology also poses a number of fascinating questions to the polymer physics community regarding modelling and control of single molecules in highly non-equilibrium environments. In particular, our model necessarily breaks down far from the life-time peak where the tug-of-war state is no longer stable.  Developing tension propagation theories appropriate for the two-pore device may be a promising new research direction\cite{sakauetp, tpreview} and would help develop a more quantitative model for two-pore tug-of-war.  Force-free translocation, while a topic of theoretical interest, has not been explored experimentally; our technology gives access to this new regime and may help motivate further work to develop more detailed understanding of the sub-diffusive physics.  Finally, developing Brownian dynamics simulation methodologies appropriate for the two-pore device will be essential in order to deduce conditions that maximize tug-of-war lifetime, minimize folding and test anti-Brownian control approaches.  Simulations could be developed to mimic the entire two-pore control process (including \emph{all} control steps i-v) and explore the effect of physical sequence-specific labels on the translocation physics.
    
   Finally, two-pore control could enhance a wide-range of existing nanopore-based technologies.  In the context of nanopore sequencing, a classic concept commercialized by Nabsys has the aim of detecting the location and number of known oligonucleotide ssDNA probes that are hybridized to a long translocating ssDNA of unknown sequence, as part of a hybridization-assisted nanopore sequencing scheme\cite{Branton:2008fr}. Detecting and locating the dsDNA regions within ssDNA should be possible with an appropriately designed dual-pore architecture.   Alternatively, positioning two biopores in sufficient proximity is technically challenging but not impossible (e.g., by alignment of substrates\cite{Lathrop:2010dk}).  Voltage field-force balancing would be facilitated by the atomic level identity of the biopore structures. Solid-state based alternatives include patterning recognition tunneling electrodes on a surface between the controlling pores\cite{Lindsay:2010dc} or as part of a multi-layered membrane\cite{Pang:2014gt}, which have sensing potential beyond nucleic acids\cite{Im:2016ec}. Beyond sequencing, the device could enable genome mapping, or detecting the presence/absence of base modifications\cite{Shim:2015ha} along with the sequence context (via PNA-PEG labeling \cite{Morin:2016js}) that is needed for clinical relevance.  Linearizing and slowing the DNA through the pore can also increase the bit library size for increased multiplexing of protein analytes.\cite{Bell:2016jr}   The two-pore control method might also be used in conjunction with ratcheting based on solid-state approaches, such as alternating membrane stacks, by lowering the forces required to hold DNA transiently immobile against external friction forces.

\section{Methods} 

\subsection{Device Fabrication}
All fabrication processes were performed at Stanford Nanofabrication Facility (SNF) and Stanford Nano Shared Facilities (SNSF). The full fabrication process is presented in sapplementary materials (Figure S10) and elsewheres\cite{Zhang:4Kttepyo, Xu:2017}. 

\subsection{Construction of a Streptavidin Labeled $\lambda$-DNA molecule}

A restriction digest map of Nt.BbvCI digested $\lambda$-DNA with expected distances between nick sites is shown in Figure S8. First 5\,$\mu$g of $\lambda$-DNA was incubated with 50 units of Nt.BbvC1 nicking endonuclease in a final volume $500\, \mu$l containing 1X cutsmart restriction enzyme (NEB) for 30 minutes. The nicking reaction was inactivated at $80\,^\circ C$ for 20 minutes. Biotin-11-dUTP was added to the heat inactivated reaction to a concentration of 50\,$\mu$M and incorporated at the nick site by nick translation with Ecoli DNA polymerase 1. A Sephadex G75 spin column was used to remove unincorporated biotin 11 dUTP and proteins. The extent of top strand nicking was evaluated by reacting $5\,\mu$l of labeled biotinylated $\lambda$-DNA with 5U of Nb.BbvC1 for 30 minutes to nicking the bottom strand. Double digested $\lambda$ fragments were resolved on 6.5\% TAE agarose electrophoresis yielding the expected restriction fragments (Fig.~S12).  $\lambda$ digestion was complete indicating that both top and bottom strand nicking reactions were complete. In order to create nanopore detectable features along the length of the $\lambda$-DNA molecule mono-streptavidin was bound to the biotinylated $\lambda$-DNA using a 1:1 ratio with respect to the concentration of NtBbvC1 nicking sites. Restriction enzyme digestion maps were generated using NEB cutter \cite{Vincze:2003aa}.

\subsection{Nanopore experiments}
The channels and common chamber were first filled with buffer (23$^\circ$C, 10\,mM Tris, 1\,mM EDTA, pH 8.8, 1\,M LiCl) using pipette by capillary force. Following testing the baselines of the two pores for noise quality (RMS noise at 30 kHz below 20\,pA), channel 1 buffer was replaced with the target reagent in buffer. Fluid replacement required 0.1\,bar pressure (nitrogen, Praxair with HARRIS connector), applied for 20 s to ensure adequate replacement. 20\,pM $\lambda$-DNA was universally used. A dual-channel voltage-clamp amplifier (MultiClamp 700B, Molecular Devices, Sunnyvale, CA) was used to apply transmembrane voltage and measure ionic current, with the 4-pole Bessel filter set at 30 kHz. A digitizer (Digidata 1440A, Molecular Devices) stored data sampled at 250 kHz. Control experiments used logic as explained in the main text, and programmed in Labview (2017) and executed on an FPGA (National Instruments model PCIe-7851).

\subsection{Data analysis}
All numerical analysis and data processing was done offline using custom code written in Matlab (2017, The MathWorks).  Single pore events were detected as described elsewhere \cite{Morin:2016js}. During co-capture events, the FPGA output for triggering $V_1$ off and reversal was recorded, and these were used to identify candidate co-capture event start times. The end of co-capture events were quantitated by the return of either current to within 1 standard deviation of the prior open pore signal mean. Failed co-capture events were also quantitated as detailed in Supplemental Sec.~5. For the streptavidin labeled $\lambda$-DNA experiments, the presence of tag sub-events within co-captured events are detected if any sample falls below the mean of the entire event minus 5 times the standard deviation ($\sigma$) of the open channel signal, with $\sigma$ computed using 1000 event-free samples prior to the onset of the co-capture event.  All other aspects of analysis of tagged events is detailed in Supplemental Sec.~7.

\begin{acknowledgement}
This work done in Santa Cruz was performed and financially supported by Two Pore Guys, Inc (2PG). For work related to device fabrication: Philip Zimny and Jen-Kan Yu. For work related to programming the FPGA: Brett Gyarfas.  The work on the McGill side was supported by funding from Two-Pore Guys and the Natural Sciences and Engineering Research Council of Canada (NSERC) Discovery Grants Program (Grant No. RGPIN 386212).   MS used in this experiment is obtained courtesy of the Howarth lab\cite{fairhead2014plug}.
\end{acknowledgement}

\section*{Competing Financial Interests Statement}
The authors declare competing financial interests: X.Liu, R.Nagel and W.B.Dunbar are employees of Two Pore Guys, Inc.~(2PG), which has exclusively licensed the dual pore device patent from the University of California, Santa Cruz, for commercialization purposes.

\bibliography{refs}

\end{document}